# The perspective of fluid flow behavior of respiratory droplets and aerosols through the facemasks in context of SARS-CoV-2

Sanjay Kumar and Heow Pueh Lee (李孝培)

Department of Mechanical Engineering, National University of Singapore, 9 Engineering Drive 1, Singapore 117575, Singapore
Author to whom correspondence should be addressed: mpesanj@nus.edu.sg; mpeleehp@nus.edu.sg

**Abstract**
In the unfortunate event of current ongoing pandemic COVID-19, where vaccination development is still at the initial stage, several preventive control measures such as social distancing, hand-hygiene, and personal protective equipment have been recommended by health professionals and organizations. Among them, the safe wearing of facemasks has played a vital role in reducing the likelihood and severity of infectious respiratory disease transmission. The reported research in facemasks has covered many of their material types, fabrication techniques, mechanism characterization, and application aspects. However, in more recent times, the focus has shifted towards the theoretical investigations of fluid flow mechanisms involved in the virus-laden particles prevention by facemasks. This exciting research domain aims to address the complex fluid transport that led to designing a facemask with a better performance. This review paper discusses the recent updates on fluid flow dynamics through the facemasks. Key design aspects such as thermal comfort and flow resistance are discussed. Furthermore, the recent progress in the investigations on the efficacy of facemasks for prevention of COVID 19 spread and the impact of wearing facemasks are presented. Finally, the potential research directions for analyzing the fluid flow behavior are highlighted.

Keywords: Facemasks; Fluid dynamics; SARS-CoV-2; COVID-19; Droplets; Aerosols

## I. Introduction

The person-to-person transmission of infectious respiratory diseases occurs primarily due to the transportation of virus-laden fluid particles from the infected person. The contagious fluid particles originate from the respiratory tract of the person and are expelled from the nose and the mouth during breathing, talking, singing, sneezing, and coughing.[1-3] These particles have been broadly classified into two types: aerosols (aerodynamic particle size < 5 μm) and droplets (aerodynamic particle size ≥ 5 μm-10 μm).[4-6] The finding indicated that the transmission phenomena of these expelled virus particles by patients would be dependent on droplet sizes. Larger respiratory droplets, once expelled from the mouth or nose, undergo gravitational settling before evaporation; in contrast, the smaller droplets particles evaporate faster than they settle, subsequently forming of the aerosolized droplet nuclei that can be suspended in the air for prolonged periods and travel in air over long distances. The research studies have revealed that the severe acute respiratory syndrome (SARS) epidemic in 2003 and the current global pandemic of coronavirus disease 2019 (COVID-19) predominantly transmitted by contact or through the airborne route.[7-10] Several preventive strategies such as safe distancing, contact tracing, isolation of the infected person, hand hygiene, and facemasks have been widely employed against the rapid spread of these diseases.[11-14] Among them, the use of the facemasks have proven to be one of the most effective protective measures against the airborne virus transmission.[15-20] The research suggested that face coverings could essentially reduce the forward distance traveled by a virus-laden droplet, and thus has a great potential to provide personal protection against airborne infection.[21,22] Recently, the World Health Organization WHO has recommended using facemasks for initial control of COVID-19 spread.[23]

In general, facemasks fall in the category of respiratory protection equipment (RPE) whose primary function is to protect the wearer from airborne viruses and contaminated fluids. There are various RPE types, ranging from simple homemade reusable cloth-based masks to surgical facemasks and N95 respirators to self-contained breathing apparatus.[18,24-27] Different types of masks provide different levels of protection to the wearer. Surgical facemasks are loose-fitting, fluid-resistant, single-time use, and disposable, designed to cover the mouth and nose. These masks are fluid resistant and intended for reducing the emission of large respiratory droplets released during coughing and sneezing.[28,29] However, there is a possibility of leakage around these facemask's edge during the inhaling and exhaling process. Such a dynamic leakage allows the direct contact of fluid droplets from the outside air to the wearer and vice-versa. Such respiratory masks may also not provide adequate protection against extremely fine aerosolized particles, droplets, and nuclei.[30]

For efficient trapping of droplets, the facemask filters should contain microscopic pores; however, the minute-sized pores prevent air ventilation, which creates an uncomfortable situation for the wearer. Hence, a better tradeoff between the pore sizes and the breathability is desirable for the suitable facemasks. Some mask types come with inbuilt respirators such as filtering facepiece respirator, P100 respirator/gas mask, self-contained breathing apparatus, full face respirator, and KN95 respirators provide better breathability of the users. The name designation 'N95' in the N95 respirators refers to the filtration of 0.3 µm sized particles with 95% efficiency.[31] The filtration mechanism of N95 facemasks operates on three possible principles: diffusion, inertial impaction, and electrostatic attraction. The smaller particles (<1 micron) usually get diffused and stuck on the filter's fibrous layers. Whereas particles of typically 1 micron or larger realize the inertia effect, preventing them from flowing across the fibers in the filtration layers slam into the mask layers and filtered. N95 masks are designed for single-use because of potential contamination of filter layers, resulting in rapid degradation of their filtration efficiency. However, several innovative techniques have been demonstrated for decontaminating and reusing N95 masks.[32,33] Some recent N95 respirator masks are fabricated using the electrocharged polymers or electrospun nanofibers.[34,35] These materials have intrinsic electrostatic properties to attract the small-to-large oppositely charged particles, which help in better filtration of small-size particle transmission.[36,37]

Because of the ongoing COVID-19 pandemic, a significant demand for facemasks has been reported worldwide, while stimulating research about their efficacy for filtering expelled droplets from the mouth and nose of the infected person. In this regard, considerable efforts have been made in the past for the evaluation of facemasks performance. The quantitative performance of the facemasks has been typically characterized by evaluating the filtration efficiency (FE) and the total inward leakage (TIL).[38-41] The filtration efficiency refers to the percentage of blocked particles by the tightly-fitted facemasks. The filtration efficiency can be calculated as $FE = \left(1 - (C_u/C_d)\right) \times 100\%$, where $C_0, C_i$ are the particle count in the upstream feed prior to filtration and in the downstream filtrate, respectively. TIL is defined as the percentage of particles entering the mask through both the filter and the leakage between mask and face. The total inward leakage is calculated by dividing the particle concentrations on the outside and inside the facemasks. The protection factor of the facemasks can be determined from the expression; PF

= 1/TIL. Higher PF value of the masks perform better in virus transmission control.[16] Furthermore, the fluid penetration resistance performance of the facemasks have been evaluated as per the ASTM F1862 /F1862M – 17 standards.[42,43] However, this test method does not evaluate facemasks' performance for airborne exposure pathways or in the prevention of the penetration of aerosolized fluids deposited on the facemask. In recent times, some qualitative analysis has been demonstrated for the rapid design characterization of facemasks.[44]

While these experimental studies are essential for the broad characterization and design evaluation of respiratory facemasks, further theoretical and numerical methods and algorithm-based investigations provide a better insight into the facemask's fluid flow dynamics and the droplet leakage through the facemask openings. If the facemask is donned for a prolonged period, the captured fluid vapor on the filter surface may reduce the filtration efficiency. This saturation effect of the facemasks has been usually neglected in the experimental studies. To involve these factors, an alternative approach, the computational fluid dynamics (CFD) method, can be invaluable for understanding the fluid-particle flow behavior through the facemasks. The fluid dynamics based numerical techniques have gained momentum in the field of the facemask research domain. The computational fluid flow models have shown their potentials in an improved prediction of the spreading of respiratory virus-laden droplets and aerosols, sensitive to the ambient environment, and crucial to the public health responses.

This review paper focuses on the fluid flow aspects of the facemasks and their efficacy in virus transmission control. Following a brief introduction to the respiratory infectious diseases and their control strategies (section I), the respiratory droplet transportation mechanisms in conjunction with the possible governing equations required for estimating the transport phenomena have been presented in section II. Then, the droplet transport behavior through the facemasks has been described in section III. Key design aspects for the facemasks have been explained in section IV. Section V covered the recent progress in investigating the efficacy of facemasks for preventing virus spread. The impact of using the facemasks have been discussed in section VI. The concluding remarks and a brief outlook for future research directions are summarized in section VII.

## II. Respiratory droplet transport governing equations

During the sneezing or coughing process, the dispersion of saliva droplets or aerosols from the mouth to the ambient, and eventually on the floor accomplish in several stages. The complete transmission cycle involves complex flow phenomena, ranging from air–mucous interaction, breaking of droplets, turbulent conical jets, droplet evaporation and deposition, flow-induced particle dispersion, and sedimentation.[45] After exhalation from the mouth or nose, the saliva droplet movement is initially led by the inertia force, followed by the formation of a conical jet (vortical flow) near the mouth. Once the droplets are expelled from the mouth, the inertia force gradually decreases, and other forces like gravity control the dispersion of larger size droplets, while drag and Brownian forces control the smaller size droplets. After traveling up to a particular distance, these virus-laden droplets settle down on the floor.[46]

Thus, there are two major possible pathways for the respiratory virus transmission: airborne inhalation of smaller droplets, which are suspended in ambient air for a more extended period and carrying to the longer distance, and contact (direct or indirect between people and with contaminated surfaces) of large size droplets.[47] The fluid flow behavior of these droplets has been modeled using two different phases: continuous phase for the small size droplet nuclei and discrete phase for large size droplets.

### A. Continuous phase

The fluid flow is governed by the Navier-Stokes and mass transfer equations which are as follows.

Continuity:
$$\nabla \cdot \vec{u} = 0 \tag{1}$$

Momentum:
$$\rho \frac{\partial \vec{u}}{\partial t} + \rho(\vec{u}.\nabla)\vec{u} - \mu \nabla^2 \vec{u} + \nabla p = 0 \tag{2}$$

Mass-transfer
$$\frac{\partial c}{\partial t} - \psi \Delta c + \vec{u}.\nabla c = 0 \tag{3}$$

where $\rho, t, \vec{u}, p, \psi, \mu$ denotes the density ($kgm^{-3}$), time ($s$), flow velocity ($ms^{-1}$), pressure ($Pa$), diffusion coefficient and kinetic viscosity, respectively. The conservation laws can be written in tensor form as:

$$\frac{\partial \rho}{\partial t} + \frac{\partial(\rho \overrightarrow{u_j})}{\partial x_j} = 0 \tag{4}$$

$$\frac{\partial(\rho \overrightarrow{u_j})}{\partial t} + \frac{\partial(\rho \overrightarrow{u_i}\overrightarrow{u_j})}{\partial x_j} = -\frac{\partial(p)}{\partial x_i} - \frac{\partial(\tau_{ij})}{\partial x_j} + S_f \tag{5}$$

Here, $\vec{u}$ represents the flow velocity (m/s) and $S_f$ is the source term that represents other forces such as gravity, Lorentz force, etc. which also leads to momentum accumulation. For the Newtonian fluids, there is a linear relationship between shear stress and velocity gradient. So, the viscous stress tensor can be defined by:

$$\tau_{ij} = -\mu \dot{\gamma}_{ij} = -\mu \left[\left(\frac{\partial(\overrightarrow{u_i})}{\partial x_j} + \frac{\partial(\overrightarrow{u_j})}{\partial x_i}\right)\right] \tag{6}$$

In the overall vector form of the constitutive equation,

$$\tau_{ij} = -\mu\left(\nabla\vec{u} + \nabla\vec{u}^T - \frac{2}{3}\nabla\vec{u}\,I\right) \tag{7}$$

where $^T$ denotes the transpose of the second velocity gradient outer product. For a Newtonian fluid with constant $\mu$ and $\rho$, the momentum equation can be rewritten as:

$$\rho\frac{\partial(\vec{u})}{\partial t} = -\nabla p - \mu\left(\nabla^2\vec{u} + \nabla.(\nabla\vec{u}^T) - \frac{2}{3}\nabla.(\nabla\vec{u}\,I)\right) + S_f \tag{8}$$

Moreover, the exhaled fluid jet may contain cough droplets combined with the environmental wind, generating a complex laminar-to-turbulent flow field. The turbulent kinetic energy (K) of the droplets can be obtained by solving the "one equation eddy-viscosity model" (OEEVM) subgrid-scale (SGS):[46,48]

$$\partial.(\bar{\rho}k) + \nabla.(\bar{\rho}ku) = -\tau_{ij}.\dot{\gamma}_{ij} + \nabla.(\mu_k \nabla k) + \bar{\rho}\varepsilon \tag{9}$$

$$\varepsilon = c_\varepsilon k^{\frac{3}{2}}/\Delta \tag{10}$$

The turbulent viscosity is caculated from

$$\mu_k = c_k \bar{\rho} \Delta \sqrt{k} \tag{11}$$

Also, the fluctuation velocity component for the laminar-to-turbulent airflow field can be predicted by the Reynolds-averaged Navier–Stokes equations (RANS) model.

$$\dot{u}_i = f_i \xi_i \sqrt{\frac{2}{3} k} \tag{12}$$

Where $\xi_i$ are the damping factors to reflect the anisotropic magnitude of the fluctuation velocity in the near-wall region. These are the random numbers from the standard normal distribution.

The cough spreading phenomena can be predicted by solving the diffusion equation (3) in conjunction with some source and sink terms. Vuorinen et al.[49] developed diffusion-based Monte-Carlo models to realize a transmission phenomenon via inhalation of aerosols in the ambient flow field. The source and sink terms have been included in conjunction with Eqn. (3). The source term represented the transient location of the infected persons while the sink term has been used for the ventilation surface. The developed models were capable of predicting the aerosol dispersions at more realistic locations like generic public place and supermarkets where cough may release from the walking person.

### B. Discrete phase (cough droplets transport and size change dynamics)

For the droplets with the high droplet-to-air density ratio, the droplet trajectories have been predicted by solving a series of translation equations (Lagrangian approach) of the discrete phase with the assumptions of stationary droplets and limited thermophoresis. Continuous dispersion of saliva droplets throughout the computational domain has been considered in the computations. Also, some basic parameters like velocity, mass, and position of each droplet have been computed at every time step. The translational equation for the saliva micro-droplet trajectory is given by,[46]

$$m_p \frac{\partial \vec{u}_P}{\partial t} = \vec{F}_D + \vec{F}_g + \vec{F}_L + \vec{F}_M \tag{13}$$

$$= \frac{3}{4} C_D \frac{\rho_f}{\rho_d} \frac{m_d}{2R_d} |\vec{u}_f - \vec{u}_d|(\vec{u}_f - \vec{u}_d) + (\rho_d - \rho_f)V_d \vec{g} + V_d \nabla P + \frac{\rho_f V_d}{2} \frac{\partial(\vec{u}_f - \vec{u}_d)}{\partial t}$$

where $\vec{F}_D, \vec{F}_g, \vec{F}_L, \vec{F}_{BM}$ are the Stokes drag force, gravity, lift or buoyancy force, and Brownian motion-induced force, respectively. Also, $m_d, R_d, V_d, \rho_d, \vec{u}_d$ are the mass, radius, volume, density, and velocity vector of the saliva droplets, respectively. $\rho_f, \vec{u}_f$ are the fluid density and the fluid velocity vector, respectively. The drag coefficient values depend on the droplet's Reynolds number $Re_d$ and can be calculated from,

$$C_D = \begin{cases} 24/Re_d & if\ Re_d < 1 \\ (24/Re_d)(1 + 0.5\ Re_d^{0.687}) & if\ 1 \leq Re_d \leq 1000 \\ 0.44 & if\ Re_d > 1000 \end{cases} \tag{14}$$

Here, $Re_d = \frac{2R_d|\vec{u}_f - \vec{u}_d|\rho_f}{\mu_f}$. In above expressions, the droplet distribution is an important factor as their size decides the travel path distance, and eventually the infection risk.[50] So, for coughing simulation the droplet breakup approach is used. Pendar and Páscoa[46] used Rosin–Rammler breakup approach in their coughing simulation work Which is expressed as:

$$f_r = \frac{qr^{q-1}}{\bar{r}^q} \exp\left[-\left(\frac{r}{\bar{r}}\right)^q\right] \quad (15)$$

Where $q$ and $\bar{r}$ are the exponential factors and average radius of the droplet, respectively. These parameters are based on the saliva flow rate.

Recently, several studies have attempted to understand the dynamics of droplet formation and transport. Cummins et al.[51] investigated the dispersion of spherical droplets in the presence of a source–sink pair flow field. The Maxey–Riley equation was used to describe the finite-sized spherical particle motion in an ambient fluid flow. The presented non-dimensional mathematical models were based on the Newton's second law of motion in which the forces acting on the particle involved the gravity force, the drag force, an added mass force, the force due to the undisturbed flow, and a Basset–Boussinesq history term. The analytical results suggested that droplets with a smaller size (<75 µm) moved a greater distance because of gravity's smaller impact. In comparison, the larger size droplets (>400 µm) traveled a relatively long distance before getting pulled into the sink by their more considerable inertia. However, the dispersion of intermediate size droplets (75 µm - 400 µm) was found to be complicated under the influence of both drag and gravity forces. Busco et al.[52] used the computational fluid dynamics approach to predict droplets and aerosols spread. The biomechanics of a human sneeze, including complex muscle contractions and relaxations, were included in the simulation by imposing a momentum source term to the coupled Eulerian-Lagrangian momentum equations (13). The instantaneous magnitude of the sneezing momentum source term has been defined as $|s(t)| = p(t)/L$, where p(t) is the experimental pressure signal, and L is the characteristic equivalent length of the human upper-respiratory system ducts. The experimental results validated the developed model for the estimation of droplets and aerosols spreads.

Das et al.[53] investigated the airborne virus transmission through sneezed and coughed droplets and aerosols. The ejected droplet motions were estimated both for still and flowing air conditions by solving the Langevin differential equation using Monte-Carlo numerical method. The Langevin equations for the transport of the droplets of mass (M) in the still air is given as,

$$\frac{dr_i}{dt} = v_i \quad (16)$$

$$M\frac{dv_i}{dt} = -\lambda v_i + \xi(t) + F_g \quad (17)$$

where $dr_i$ and $dv_i$ are the coordinate and velocity shift in each discrete time step $dt$, respectively, and $i$ stands for the Cartesian components of the position and velocity vectors. The first term in the right-hand side of Eq.(17) represents the dissipative force. The second term stands for the diffusive (stochastic) force where ξ(t) that is regulated by the diffusion coefficient D. $F_g$ is the gravitation force term acting on a droplet of mass M. In the expression, the value of the drag coefficients $\lambda$ is obtained using the Stokes formula, $\lambda = 6\pi\eta R$, here $R$ is the droplet radius and $\eta$ is viscosity. The diffusion coefficient D is obtained from the Einstein relation, $D = K_B T \lambda$, where $K_B = 1.38 \times 10^{-23} J/K$ is the Boltzmann constant and T is the temperature in Kelvin. As shown, the Langevin differential equations contain a stochastic source term (diffusive force), which is usually ignored in the Eulerian-Lagrangian approach. Also, environmental factors such as temperature, humidity, and airflow rate, which could influence the air droplet dynamics, were included. The results revealed that the small droplets travel a larger distance and remain suspended in the air for a longer time under the influence of airflow, supporting the mandatory use of facemasks to prevent the virus.

Vadivukkarasan et al.[54] experimentally investigated the Breakup morphology of expelled respiratory liquid. It was revealed that the droplet formation from the ejected fluid during coughing or sneezing occurred due to three possible mechanisms: Kelvin–Helmholtz (K–H) instability, Rayleigh–Taylor (R–T) instability, and Plateau–Rayleigh (P–R) instability in sequence. The flapping of the expelled liquid sheet was the result of the K–H mechanism, and the ligaments formed on the edge of the rim appeared due to the R–T mechanism, and finally, the hanging droplet fragmentation was the result of the P–R instability.

### C. Droplet evaporation

Droplet evaporation is one of the crucial factors that affect transmission phenomena. The evaporation rate of the droplets depends on the difference between the saturated vapor pressure of the fluid droplet surface and the vapor pressure of the surrounding air (ambient temperature and humidity).[55] The other factors, such as the mass-diffusion coefficient and the relative velocity between the droplet and surrounding gas, influence the evaporation rate. The non-dimensional parameters such as Reynolds, Nusselt, and Sherwood numbers govern the droplet evaporation phenomena.[56] Moreover, the condensation and evaporation effects between the ambient water vapors and the water liquid in cough droplets can be considered by solving the mass and energy balance for each droplet.[57]

Mass balance:
$$\frac{dm_d}{dt} = -\sum_{e=1}^{k}\int_{surf} n_e dA \approx \sum_{e=1}^{k}(\bar{n}_e \, dA) \tag{18}$$

Energy balance:
$$\sum_{i=1}^{m} m_{d,i}\, c_{d,i} \,.\, \Delta T = \pi d_d \lambda_g \, Nu \, (T_a - T_d) - \sum_{e=1}^{k}\iint_d n_e \, L_e \, dA \tag{19}$$

where $n_e$ is the average mass flux of evaporable component $e$ on the surface that can be expressed as:

$$n_e = \frac{\rho_g \, Sh \, \widetilde{D}_e C_m}{d_d} \ln\frac{1 - Y_{e,\infty}}{1 - Y_{e,surf}} \tag{20}$$

Where $\rho_g$ is the density of the ambient air, $Y_{e,surf}$ and $Y_{e,\infty}$ are the mass fractions of evaporable component $e$ on the droplet surface and in the gas phase far from the droplets, respectively. $Sh$ is the Sherwood number.

$$Sh = \sqrt{1 + Re_d.Sc} \,.\, \max[1, Re_d^{0.077}] \tag{21}$$

Where $Sc = \mu/\rho D_e$ is the Schmidt number, and $D_e$ is the mass diffusivity of component $e$. The Nusselt number is calculated as

$$Nu = (1 + Re_d . Pr)^{0.33} \max[1, Re_d^{0.077}]$$

(22)

Here, $Pr$ is the Prandtl number. A detailed explanation of other variables has given in the previous published article by Feng et al.[58]

Several other researchers have studied the flow behavior of evaporating droplets. Recently, Weiss et al.[59] investigated the clustering and evaporation of droplets using the gas phase and droplet coupling equations. The evaporation of droplets and spreading of vapors into the ambient condition were mostly governed by few parameters: the Reynolds number, which is related to the shear rate, the Stokes number, and the mass loading, which is the ratio between the mass of the liquid to the gas phase.[60] The results suggested that the clustering and evaporation of droplets are primarily affected by the mass loading and Stokes number while the Taylor-scale Reynolds number was small. When the mass loadings decreased, and the Stokes number increased, the droplets dispersed more evenly with a faster evaporation rate. Chaudhuri et al.[61] presented a chemical reaction mechanism based collision rate model for prediction of the growth rate of the infected population for the early phases of a Covid-19 like pandemic. Besides, they developed a theoretical model for the aerodynamics of respiratory droplets by considering the evaporation characteristics of levitated droplets. The evolution of the droplets was characterized by a complex interaction of aerodynamics, evaporation thermodynamics, and crystallization kinetics. The fidelity of proposed model was further confirmed by the experimentation.

### III. Respiratory droplet transmission through the facemasks

Respiratory droplet transmission is considered critical for the rapid spread and continued circulation of viruses in humans. In recent years, the respiratory droplets flow behavior through the facemasks has typically well-predicted using the computational fluid dynamics (CFD) techniques.[21] The Navier-Stokes equations have been used as basic governing equations to solve the velocity field in a multi-dimensional computational domain. These equations have been used for the analytical assessment of the respiratory performance of the facemasks and other respirators. Dbouk and Drikakis[21] performed the fluid dynamics analysis of the respiratory droplets transmission through and around a facemask filter. The compressible Reynolds-averaged Navier–Stokes equations and the k–ω turbulence model were employed. Zhang etl.[62] analytically investigated the carbon dioxide $CO_2$ transportation performance inside the ventilator mask. The 3D model of the ventilator mask is shown in **FIG. 1**a. Classical Navier-stokes theorem and mass-transport equations were used to estimate the $CO_2$ residual concentrations below the nostrils. The governing equations were solved using the finite element solver ANSYS fluent 15.0 software. The following governing equations were used in the simulation; (i) at the entrance of the ventilator mask, the inlet pressure = $0.98 \times 10^3 Pa$, the average concentration of $CO_2$ = 0.03%. (ii) at the exhaust holes: outlet pressure = $0\ Pa$, (iii) Inlet boundary condition at the nostrils: the averaged velocity $\bar{u} = 6 \times \sin\left(\frac{\pi}{2}t\right)$, expiratory phase time t = 0~2.0 s, inspiratory phase time t = 2.0~4.0 s, and the averaged concentration of $CO_2$ excreted from the nostrils was set as 4%. The airflow inside the ventilator mask was considered to be turbulent flow. **FIG. 1**b shows the distribution of the average residual $CO_2$ concentration inside the ventilator mask varying with time during a complete respiratory cycle. As shown from the curve, initially, the $CO_2$ concentration increased with the increasing exhaled air and reaches the peak value of 3.65 %, and then it declined gradually with the decrease of the exhaled air and reaches down to the value of 1.8% at the end time of expiratory cycle. Based on these results, the ventilator mask was redesigned by changing the exhaust hole to the bottom side and the local residual $CO_2$ concentration was decrease to 0.7%.

Bates et al.[63] performed computational fluid dynamics simulations to access the respiratory airflow in the human upper oral airway with airway wall movement. The breathing flow rate data was acquired by imaging the breathing cycle of the participant while wearing of a size-5 anesthesia facemask (**FIG. 1**c). The air pressure drop and flow velocity were estimated by solving the Navier-stokes equations for the moving mesh vertices in the finite volume domain. The governing equations for moving mesh of the finite volume form is given by:

Continuity equation:
$$\frac{\partial}{\partial t}\int_V \rho \, d\tilde{V} + \oint \rho(\vec{u} - \vec{u_g}).d\vec{a} = 0 \tag{23}$$

Momentum equation:
$$\frac{\partial}{\partial t}\int_V \rho \, d\tilde{V} + \oint (\rho(\vec{u} - \vec{u_g}) \otimes \vec{u}).d\vec{a} = -\oint p\mathbf{I}.d\vec{a} + \oint \mathbf{T}.d\vec{a} \tag{24}$$

where $t$ is time, $V$ is the volume of each cell in the mesh, $\rho$ is the air density, $\vec{u}$ is the air flow rate, $\vec{u_g}$ is the mesh velocity as calculated from the mesh displacement for each control points, $\vec{a}$ is a vector representing the surface of each mesh cell, $\mathbf{I}$ is the identity matrix, and $\mathbf{T}$ is the viscous stress tensor. These equations were solved using the large eddy simulation (LES) techniques. The instantaneous air flow resistance was calculated as the pressure loss between two locations divided by the air flow rate through them. **FIG. 1**d shows the estimated airflow resistance through several different regions of the extra thoracic airway during the complete breathing cycle.

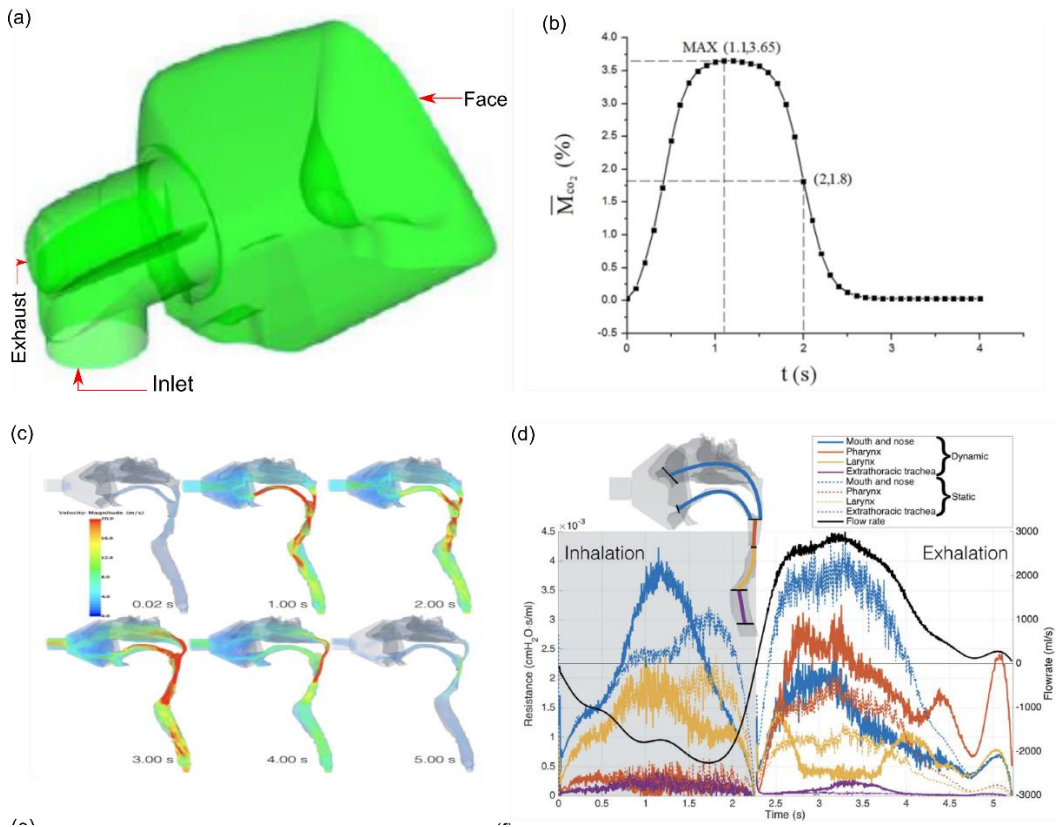

**FIG. 1. (a)** The 3D schematics of the ventilator mask integrated with the volunteer's face. **(b)** distribution of the averaged residual $CO_2$ concentration inside the ventilator mask varying with time during a complete respiratory cycle. Reprinted with permission from Zhang et al.[62], © 2018 Tech Science Press, licensed under a Creative Commons Attribution 4.0 International License. **(c)** The surface of the airway model at six instants through the breathing maneuver. The model was extended from the mask worn by the wearer. **(d)** The resistance to airflow through the breath. The colored lines represent the resistance (left axis) through each of the regions between the planes shown in the inset (top left). The solid lines show the resistance in the moving wall simulation, while the dashed lines show the resistance in the same regions in the static geometry. The black curve shows the flow rate throughout the breath (right axis). Reprinted with permission from Bates et al.[63], © 2017 Elsevier Ltd.

The aerosol droplets transmission phenomena through the facemasks have also been investigated analytically. The facemask leakage factor has been considered in the analytical models. Lei et al.[64] predicted the fluid leakage between an N95 filtering facepiece respirator (FFR) and a headform using the computational fluid dynamics (CFD) simulation approach. The mass flow rate at the faceseal and through the filter medium was calculated under three different boundary conditions: varying breathing velocity, varying viscous resistance coefficients of the filter, and the freestream air flows. The filter-to-faceseal leakage (FTFL) ratio for the respirator was obtained by dividing the mass flow rate through the filter medium and the faceseal leakage. A higher FTFL ratio refers to the higher percentage of airflow passing through the filter medium than the faceseal leakage. The results revealed the nonlinear increase in the FTFL ratio with increasing breathing velocity values and decreasing the filter viscous resistance coefficient values. Furthermore, the freestream flow had limited influence on the airflow inside the respirator resulting in nonsignificant variations on the FTFL ratio. Perić et al.[65] investigated the one-dimensional fluid dynamics of the facemasks using analytical and numerical computations. For simplifying the problem, a hemi-spherical geometry was selected for the analysis. **FIG. 2**a shows the schematic representation of a hemi spherical facemasks with possible fluid flow directions. When inhaling or exhaling, the total (volumetric) flow rate of the fluid through the nose $F_t$ is given as per mass conservation laws.

$$F_t = F_g + F_m \tag{25}$$

The volumetric flow rate can be calculated as $F_i = u_i S_i$, where $u_i$ is the average flow velocity and $S_i$ is the cross-sectional area. In the expression, the subscript $i$ denotes the nose ($t$), airgap ($g$) and mask filter ($m$). Moreover, the fluid flow through the gap is considered as fully developed laminar Poiseuille flow because the average gap velocity is estimated to be below the critical velocity $u_{crit} \approx \frac{vRe_{crit}}{D_h}$, where $D_h$ is the hydraulic diameter, and $Re_{crit}$ is the Reynolds number. Also, fluid passes through the gap between the face and facemasks, a pressure drop can be observed at inlet, inside the gap, and at outlet. The pressure drop at gap inlet and outlet is given as

$$\Delta P_{g,1} = \frac{u_t}{|u_t|} \xi \frac{\rho}{2} u_g^2 \tag{26}$$

The pressure drop inside the gap is given by

$$\Delta P_{g,2} = \frac{12\mu L_g}{H_g^2} u_g \tag{27}$$

The total pressure drop through the gap must be equal to the pressure drop through the mask filter-piece.

$$\Delta P_m = \Delta P_{g,1} + \Delta P_{g,2} \tag{28}$$

Where $\Delta P_m = C_m \rho_f u_m$, $C_m$ is the viscous porous resistance of the mask filter material, $\rho_f$ is the fluid density and $u_m$ is the average flow velocity through the mask that can be computed from the expression $F_m = u_m S_m$. The Eqns. (25)-(28) are used to theoretical estimation of fluid flow behavior.

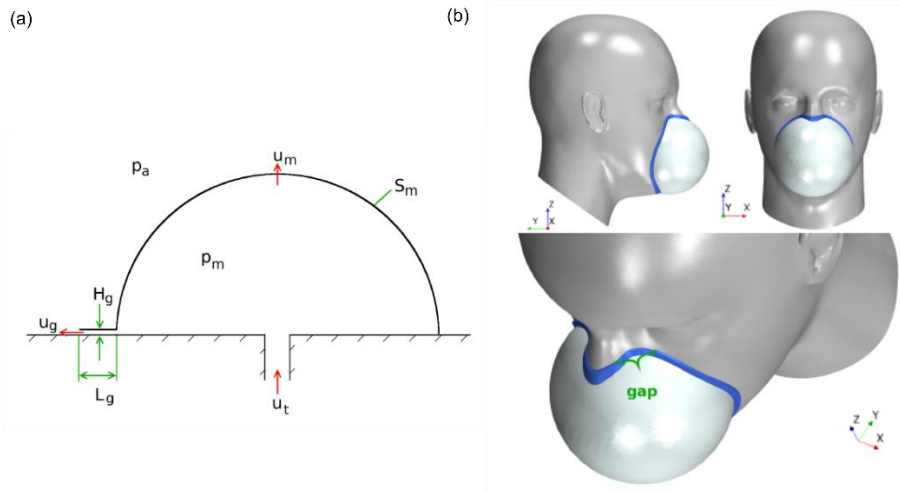

**FIG. 2. (a)** Schematic geometry for airflow through a generic hemi-spherical facemask with gap height $H_g$ and gap length $L_g$ over a width $B_g$ along its perimeter. $u_t$, $u_m$ and $u_g$ denote average airflow velocities through nose cross-sectional area $S_t$, mask filter surface $S_m$ and gap cross-sectional area $S_g$. $p_m$ and $p_a$ signify the pressure inside and outside (atmospheric) of the mask. **(b)** 3D representation of facemasks with wearer showing the possible region for leakage. Reprinted with permission from Perić and Perić[65], © 2020 arXiv.org.

## IV. Key design aspects

**Thermal comfort**

Thermal comfort is an essential aspect of a facemask as it may affect the compliance of the use of facemask during summer or in tropical countries. There were reported incidence of skin rashes, increased heat stress, sweating, and discomfort due to prolonged wearing of a facemask in hot and humid conditions.[66] To improve the thermal comfort level of facemasks, researchers have developed some unique facemasks by using the nanocomposites. Polymer-based nanofibers with large surface area-to volume ratio have shown great potential for use in facemasks to achieve both high filtration efficiency and sufficient air permeability.[67-69] Yang et al.[70] presented a design of the nanofiber based facemasks for a better thermal comfort of the user. The facemask was made of hybrid nanocomposites containing electrospun nylon-6 nanofibers on top of needle-punched nanoporous polyethylene (nanoPE) substrate. While nanofibers with strong particulate matter (PM) adhesion properties ensured high PM capture efficiency (99.6% for PM2.5) with low pressure drop, and nanoPE substrate with high infrared (IR) transparency (92.1%, weighted based on human body radiation) resulted in effective radiative cooling. **FIG. 3**(a-c) show the schematic, photographs, and scanning electron micrographs of the proposed hybrid nanofiber-based facemask. The comparative PM capture efficiency and air permeability results have demonstrated the superiority of presented facemask over the commercial masks (**FIG. 3**d, e). Moreover, the thermal image revealed that the fiber/nanoPE facemasks had high transparency to the human body radiation (cooling effect). In contrast, the commercial facemasks blocked a large portion of it. They further modified the nanoPE substrate with Ag coating and demonstrated that fiber/Ag/nanoPE had a warming effect.

Zhang et al.[71] reported the use of an active ventilation fan to reduce the dead space temperature and $CO_2$ level. An infrared camera (IRC) method was used to elucidate the temperature distribution on the prototype FFR's outside surface and the wearer's face, surface temperature was found to be lowered notably. Both inside and outside temperature resulted from the simulation were found to be in good agreement with experimental results. However, the inward blowing fans may compromise the filtering effectiveness of the facemask. There are commercially available facemasks fitted with one-way valve for facilitating the removal of humidity and expired air within the space between the facemask and the face. However, during the covid-19 pandemic, one of the main reasons for wearing the mask is not only to protect the inhalation of virus, but also to prevent the spread of virus into the air if the wearer happens to be a carrier of the virus. If the wearer is a healthy subject, the use of a one-way valve and ventilation fan would indeed mitigate the buildup of humidity and carbon dioxide within the dead space. Zhu et al.[72] reported a three-dimensional model of normal human nasal cavity to simulate the volume of fraction of both fresh air and respired air within the nasal cavity. The model consisted of large rectangular domain outside the nasal cavity representing ambient air, human nasal cavity and partial of the pharynx. This was the first reported piece of work that modelled the details of nasal cavity instead of just the nostrils as openings for the flow simulations. The advantage for this simulation was that the flow field within the space between the nostrils and facemask could be more accurately simulated as the boundary condition could be specified away from the nostril at the pharyngeal area. Two cases were simulated. Case I refers to a human face with a N95 respirator onto human face, and case II refers to a human face without a respirator. The results showed that above 60% of inspired air was respired air in case I compared to less than 1.2% in case II. During expiration, the volume of fraction (VOF) of respired air in both cases was above 95%. The streamlines at peak inspiration were relatively smooth while entering the cavity in both cases; while at peak expiration large vortex was observed within the air space between human face and respirator in case I. For future studies, one could explore the in vivo experimental studies with the use of miniaturized and wireless sensors for monitoring not just the temperature, but also the humidity and carbon dioxide content within the space between the nostrils and the facemask. The sensors need to be small so as not to disrupt the flow fields. If a single sensor cannot be small enough for the measurement of all the three parameters, one may need to have separate sensors and repeat the experiment for the same human subject.

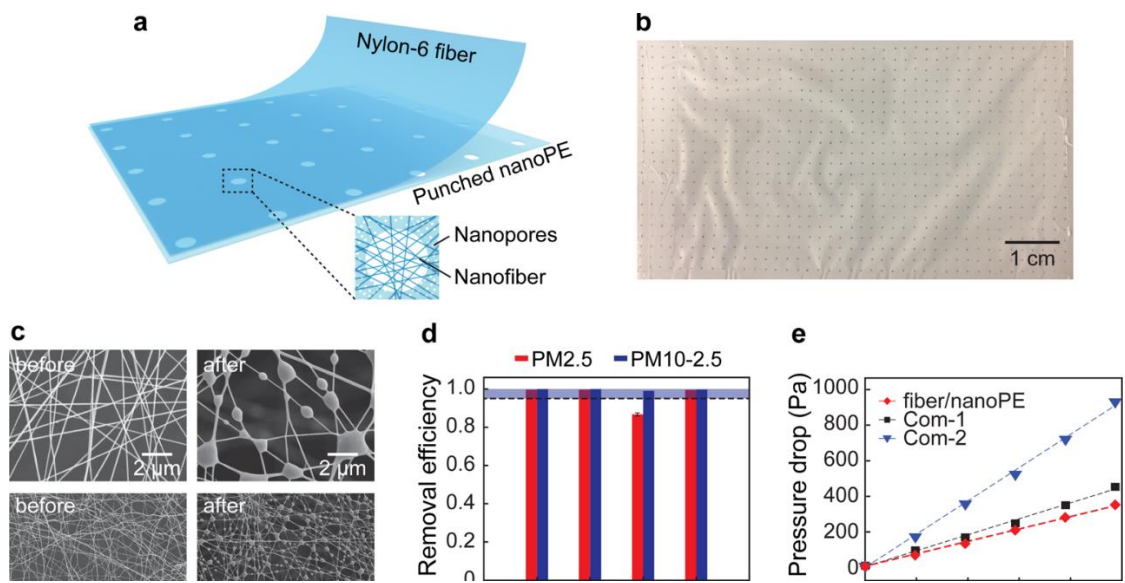

**FIG. 3. (a,b)** Schematics of proposed hybrid facemask (nanofibers/nanoPE) and its photograph. **(c)** The SEM images shows the condition of nylon-6 fibers before and after filtering the particulate matter (PM). **(d,e)** The removal efficiency of the fiber/nanoPE facemasks compared to two commercial masks, and their pressure drop spectra as a function of the wind velocity. Reprinted with permission from Yang et al.[70], ©2017 American Chemical Society.

**Flow resistance**

Another important parameter affecting the comfort of the wears is the flow resistance of the facemask. In principle, if the flow resistance is lower while maintaining the same filtering efficiency, the comfort level will be enhanced. However, the facemask's flow resistance is just an indicator and does not specify the wearer's breathing resistance. While the flow resistance could be measured using a typical setup for correlating the fluid flow rate to the pressure drop across the facemasks, the breathing resistance could only be measured using a human subject or a replica of the nasal pharyngeal system. Lee and Wang[73] presented the pioneering work of measuring the Nasal airflow resistance during inspiration and expiration using a standard rhinomanometry and nasal spirometry. A modified full-facemask was produced in-house to measure nasal resistance using N95 (3M 8210) respirators. The results showed a mean increment of 126 % and 122% in inspiratory and expiratory flow resistances, respectively, with N95 respirators. There was also an average reduction of 37% in air exchange volume with the use of N95 respirators.

The same group did a follow-up study investigating the change of human nasal functions after wearing an N95 respirator and surgical facemask.[74] The human subject study involved 87 healthy healthcare workers. Each of the volunteers attended two sessions and wore an N95 respirator in session 1 (S1) and surgical facemask in session 2 (S2) for 3 hours. The mean minimum cross-sectional area (mMCA) of the two nasal airways via acoustic rhinometry and nasal resistance via rhinomanometry was measured before and immediately after the mask. The equipment could not perform in vivo measurement with the facemask on. Rhinomanometry was repeated every 30 minutes for 1.5 hours after the removal of masks. A questionnaire was distributed to each of the volunteers during the 3 hours mask-wearing period to report subjective feelings on the discomfort level of breathing activity. Among 77 volunteers who completed both the two sessions, the mean nasal resistance immediately increased upon removing the surgical facemask and N95 respirator. The mean nasal resistance was significantly higher in S1 than S2 at 0.5 hours and 1.5hours after removing the masks ($p<0.01$). There was an increase of nasal resistance upon removal of the N95 respirator and surgical facemask, potentially due to nasal physiological changes. N95 respirator caused higher post-wearing nasal resistance than surgical facemask with different recovering routines. This was the first time that the effect of long duration wearing a facemask was objectively monitored. However, the during of three hours for the wearing of a facemask was deemed to be too short under the current Covid-19 simulations, and human subject study for a longer duration of wearing facemask should be attempted. The research could also be enhanced using miniaturized pressure, temperature, humidity, and gas sensors for in vivo monitoring of the air condition within the space between the nostrils and the facemask. Such experimental data would be useful for validating numerical models for assessing the comfort level for wearing different types of facemask. Another potential approach is to develop a replica for replacing human subject for such long duration study, similar to the use of an acoustic head for replacing human subjects in the more extended duration noise exposure study.

Zhu et al.[75] reported another investigations on effect of long duration wearing of N95 and surgical facemasks on upper airway functions. A total of 47 volunteers of National University Hospital Singapore were participated for the study. Each of the volunteers wore both N95 respirator and surgical facemask for 3 hours on two different days. During the period of mask wearing, relative airflow rates were recorded. The study revealed that the increased level of discomfort to the user with time while wearing the masks. Moreover, N95 respirator caused higher post-wearing nasal resistance than the surgical facemask with different recovering routines.

## V.     Effectiveness of facemasks for prevention of virus transmission

The current studies recognized that the airborne transmission of aerosols produced by asymptomatic individuals during speaking and breathing as a key factor leading to the spread of infectious respiratory diseases such as COVID 19.[58,76-78] However, the spread of these airborne diseases has been successfully controlled up to a certain extent by using the facemasks.[11,19,47,79-81] In the ongoing global pandemic of the COVID 19, where vaccine developments still at a phase trial stage, the respiratory protective equipment such as facemasks has proven to be a complementary countermeasure against the spread of the novel coronavirus. In this regard, several researchers have performed theoretical and experimental investigations of virus transmissibility through the facemasks and alternatives. Stutt et al.[82] developed the holistic mathematical frameworks for assessing the potential impact of facemasks in COVID 19 pandemic management. The results revealed that professional and home-made facemasks were highly efficacious to reduce exposure to respiratory infections among the public. Also, when people wear the facemasks all-time at the public places, the certain epidemiological threshold, known as the effective reproduction number, could be decreased below 1, leading to the prevention of epidemic spread. Ngonghala et al.[83] developed a parametric model for providing deeper insights into the transmission dynamics and control of COVID-19 in a community. They used the COVID 19 data from New York state and the entire US to assess the population-level impact of the various intervention strategies. The results suggested that the consistent use of facemasks could significantly reduce the effective reproduction number. The highly efficacious facemask, like surgical masks with estimated efficacy of around 70%, could lead to the eradication of the pandemic if at least 70% of the residents use such masks in public consistently. The use of low efficacy masks, such as cloth masks with an estimated efficacy of 30%, could also lead to a significant reduction of COVID-19 burden. Yan et al.[84] evaluated the effectiveness of different respiratory protective equipment in controlling infection rates in an influenza outbreak. They used a previously developed risk assessment model[85] to show N95 respirators' efficacy, low-filtration surgical mask (adult), high-filtration surgical mask (adult), high filtration pediatric mask, and low filtration pediatric mask. The study revealed that donning these masks with a 50% compliance rate resulted in a significant reduction in transmission risk, and with 80% compliance rate nearly eradicated the influenza outbreak. Prasanna Simha and Rao[86] quantitatively investigated the distance of travel of typical human coughs with and without different masks: disposable three-ply surgical masks and N95 masks. In their study, the schlieren method, a highly sensitive, non-intrusive flow imagining technique, was used to visualize the human cough flow features. The experimental statistics showed that the propagation of a viscous vortex ring mainly governed cough flow behavior. While wearing regular face masks, the cough droplets traveled approximately half the distance traveled by expelled droplets without a mask. However, N95 was found to be most effective in limiting the spread of cough droplets. Leung et al.[87] performed experimental studies to investigate the efficacy of surgical facemasks to prevent respiratory virus shedding. The surgical facemasks' efficiency was measured against the coronavirus, influenza virus and rhinovirus of two broad particle sizes, respiratory droplets (≥5μm) and aerosols (droplet nuclei with aerodynamic diameter ≤5μm). The results indicated that surgical facemasks could efficaciously prevent transmission of human coronaviruses and influenza viruses into the environment in respiratory droplets, but no significant reduction in aerosols.

Moreover, the steep rise in demand for medical facemasks during the current pandemic COVID 19 has resulted in a subsequent breakdown of the global supply chain that led to an acute shortage in the market. To mitigate this discontinuous supply chain system, scientists have put much effort into exploring alternative fabrics with sufficient filtering capacity that are readily available and affordable. Kähler and Hain[88] performed a detailed analysis of the efficacy of facemasks to prevent virus spread. In the first step, the transmission of droplets released by the mouth when breathing, speaking, and coughing was characterized. Then, the filtering capacity of the various facemasks was analyzed. The experimental results have shown that most household materials tested do not provide much protection against the virus transmission via droplets and, therefore, unsuitable as materials for protective masks. However, filtering facepiece respirators (FFR) performance-based masks such as FFP2 (Europe EN 149-2001), N95 (United

States NIOSH-42CFR84), DS2 (Japan JMHLW-Notification 214, 2018), and KN95 (China GB2626-2006) offer adequate protection, as they are only permeable to a tiny fraction of few micron-sized droplets. Konda et al.[89] evaluated the filtration efficiency of various commonly available fabrics, including cotton, silk, chiffon, flannel, various synthetics, and their combinations, which were used in the fabrication of cloth masks. The filtration performance of these fabrics was conducted by generating the aerosol particles at the cloth sample's upstream side. The aerosol particulates ranging from ~10 nm to ~ ten µm scale sizes, particularly relevant for respiratory virus transmission, were produced by commercial sodium chloride (NaCl) aerosol generator. Also, the air with a controlled airflow rate was drawn through the sample using a blower fan. The filtration efficiency $\eta_f$ of each sample was computed by measuring the particles' concentration upstream and downstream as $\eta_f = \frac{C_u - C_d}{C_u} \times 100$, where $C_u$ and $C_d$ are the mean particle concentrations per bin upstream and downstream, respectively. Moreover, the pressure drop across the facemasks and the air velocities were measured using a digital manometer and a Hot Wire anemometer. The experimental investigations revealed that the materials such as natural silk, a chiffon weave (90% polyester–10% Spandex fabric), and flannel (65% cotton–35% polyester blend) provided good electrostatic filtering of particles. Also, fabric with tighter weaves and low porosity, such as cotton sheets with high thread count, have resulted in better filtration efficiencies. For instance, a 600 TPI (thread per inch) cotton sheet can provide average filtration efficiencies of 79 ± 23% (in the 10 nm to 300 nm range) and 98.4 ± 0.2% (in the 300 nm to 6 µm range). A cotton quilt with batting provides 96 ± 2% (10 nm to 300 nm) and 96.1 ± 0.3% (300 nm to 6 µm). Surprisingly, four-layer silk (e.g., scarf) was found to be effective with an average filtration efficiency of >85% across the ten nm–6 µm particle size range. Moreover, the hybrid masks made by combinations of two or more fabric types, leveraging mechanical and electrostatic filtering, could be an effective approach for better filtration (**FIG. 4**a). Verma et al.[44] performed the qualitative investigations for assessing the effectiveness of easily available facemasks such as bandana (elastic T-shirt material, 85 threads/inch), folded handkerchief (cotton, 55 threads/inch ), stitched mask (quilting cotton, 70 threads/inch) and other commercial masks. They observed that a stitched mask made of quilting cotton was most effective, followed by the commercial mask, the folded handkerchief, and, finally, the bandana. Their observations also suggested that a higher thread count by itself is not sufficient to provide a better droplet filtration capability. The material types and fabrication techniques have a significant impact on the performance of facemasks. Davies et al.[39] examined the efficacy of homemade masks as an alternative to commercial surgical masks. Various household materials such as 100% cotton T-shirt, scarf, tea towel, pillowcase, antimicrobial pillowcase, vacuum cleaner bag, cotton mix, linen and silk were evaluated for the capacity to prevent bacterial and viral aerosols transmission. The performance of these household facemasks was compared with the standard surgical mask. The experimental outcomes showed that these homemade masks could reduce the likelihood of infection, but not efficient for the complete elimination of risks. A similar conclusion has been made in a previously published review article by Rossettie et al.[90] and Loupa et al.[91] Recently, Ho et al.[92] investigated the droplet filtration efficiency of the self-designed triple-layer cotton masks, their performance was compared with the standard medical mask. All tests were performed in two different locations; in a regular bedroom and a car with air conditioning. The particles with a size range of 20–1000 nm were taken into consideration, and the filtration efficiency was measured. Other factors like environmental conditions (temperature and relative humidity) and cough/sneeze counts per hour were measured for each measurement. The results revealed that cotton and surgical masks could significantly reduce the number of microorganisms expelled by participants with the filtration efficiency of 86.4 % and 99.9 %, respectively (**FIG. 4**b). However, the surgical mask was three times more effective in blocking transmission than the cotton mask. In a recent study, Fischer et al.[93] performed testing of 14 different facemasks or mask alternatives ranging from the kind worn by healthcare professionals to neck fleeces and knitted masks. **FIG. 4**c shows the photographs of the facemasks and alternatives considered in the investigation. A comparison was made on the dispersal of droplets from a mask wearer's breath while wearing one of the face coverings to the results of a controlled trial where their mouth was fully exposed. The study revealed that some mask types matched standard surgical masks' performance, while some mask alternatives, such as neck fleece or bandanas, offered little protection against infection. The neck fleece was found to increase the risk of disease by having a "droplet transmission fraction" of 110% (**FIG. 4**d). Besides, they demonstrated a simple optical measurement method to evaluate the efficacy of facemasks to reduce respiratory droplets transmission during regular speech. **FIG. 4**e shows the schematic of developed setup. The proposed optical system is inexpensive and easy-to-operate, even by non-experts.

Furthermore, the use of face shields has widely been used along with standard face masks. Face shields are generally made of transparent plastic sheets. They offer several advantages: comfortable to wear, easy-to-clean, clear conversations between the speakers with visible facial expressions, and reduce autoinoculation by preventing the wearer from touching their face.[94] Also, face shields prevent the user's face from the direct contact of liquid droplets. More recently, Verma et al.[95] investigated the effectiveness of the face shields and exhalation valves in the respiratory droplets transport context. They performed experimentation in an emulated coughing and sneezing environment for qualitative visualizations analysis. The results indicated that although face shields block the initial forward motion of the fluid jet, the expelled droplets can move around the visor with relative ease and spread out over a large area depending on environmental conditions. Also, for the facemasks equipped with an exhalation port, the droplets pass through the exhalation valves. Based on the observations, they opined that high-quality cloth or surgical masks perform better than the face shields and exhalation valves.

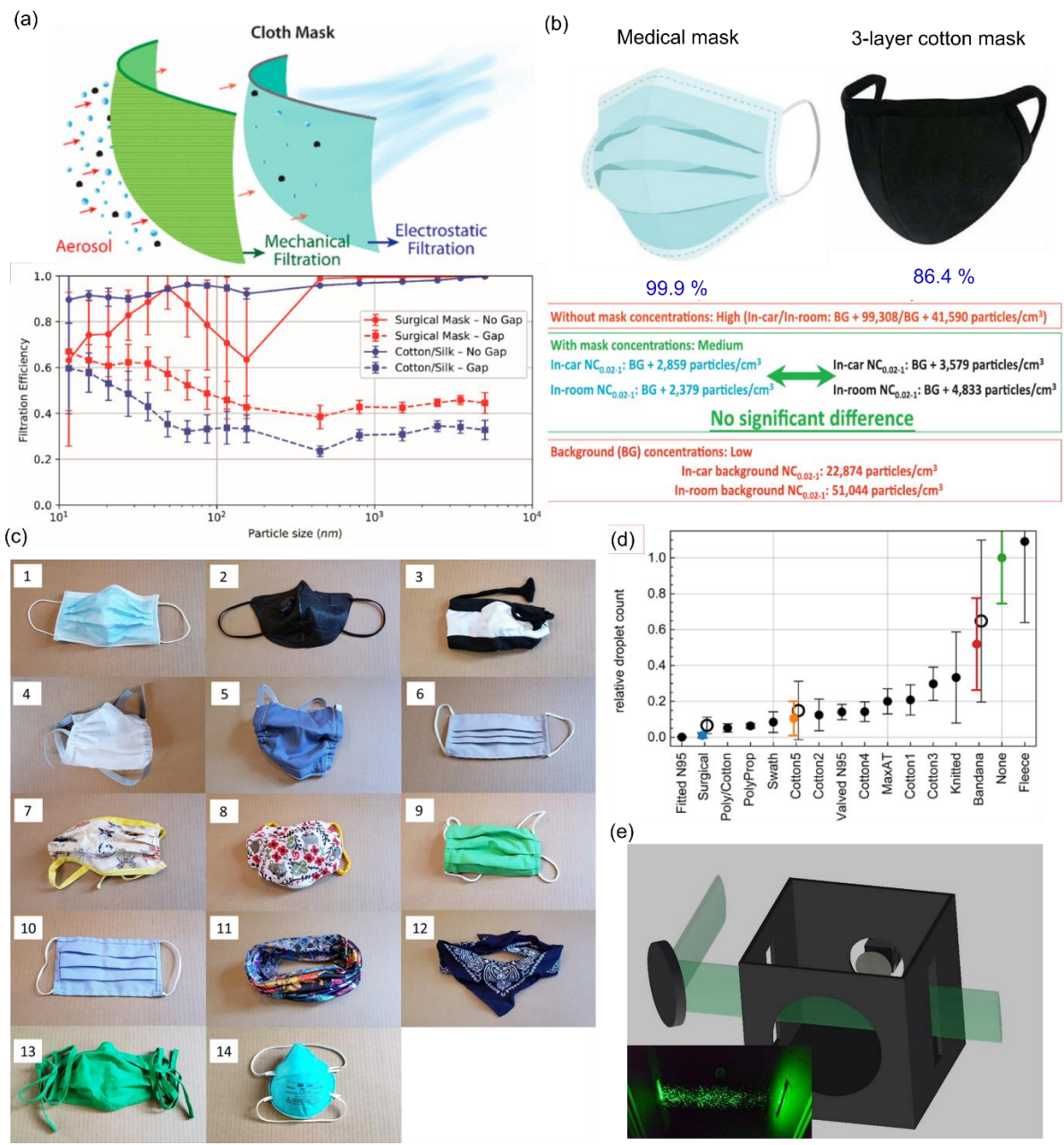

**FIG. 4. (a)** Schematic illustration of possible filtration mechanism of the hybrid cloth masks. Also, the plot shows the filtration efficiencies of a surgical mask and hybrid fabrics cotton/silk with (dashed) and without a gap (solid). The gap used was ~1% of the active mask surface area. Reprinted with permission from Konda et al.[89] ©2020 American Chemical Society. **(b)** Performance comparison between the medical masks and three-layer cotton mask. Reprinted with permission from Ho et al.[92] ©2020 Elsevier B.V. **(c)** Photographs of the facemasks under investigation. 1. Three-layer surgical mask, 2. N95 mask with exhalation valve 'Valved N95', 3. Knitted mask, 4. Double-layer polypropylene apron mask 'Polyprop', 5. Cotton-polypropylene-cotton mask 'Poly/cotton', 6. Single layer Maxima AT mask 'MaxAT', 7. Double-layer cotton- pleated style mask 'Cotton2', 8. Double-layer cotton mask- Olson style mask 'Cotton4', 9. Double -layer cotton- pleated style mask 'Cotton3', 10. Single-layer cotton- pleated style mask 'Cotton1', 11. Gaiter type neck fleece 'Fleece', 12. Double-layer bandana 'Bandana', 13. Single-layer cotton- pleated style mask 'Cotton5', 14. N95 mask no exhalation valve fitted 'Fitted N95'. **(d)** Relative droplet transmission through the corresponding facemasks. **(e)** Schematic of the experimental optical setup. Reprinted with permission from Fischer et al.[93], ©2020

## VI. Impact of using facemasks and recent designs

In the past few decades, especially post-outbreak of the severe acute respiratory syndrome (SARS) in 2003, wearing the facemasks has grown extensively. The people from Asian countries like China, Singapore, Thailand, Japan, etc. can be easily seen donning facemasks in public places. There are well proven-studies about the prevention of airborne pathogens transmission by covering the mouth and nose using the facemasks. The recently published article by Gandhi and Rutherford,[96] claimed that the universal facial masking might help reduce the severity of disease and enhance the wearer's immunity. However, prolonged use of facemasks has some side effects on human respiratory health, such as carbon dioxide builds up, drowsiness, and breathing problems because of restricted fresh airflow, and unusual heart rate.[97,98] If a facemask is donned for a longer period, the filter gets wet because of facial sweat, and vapor is formed inside the facemasks due to the breathing, resulting in clogging of particulates. Also, wearers get a false sense of security, encouraging them to spend more time in public places.[99] Other potential side-effects of facemasks wearing include skin irritation, uncomfortable feeling due to the arrival of exhaled air into the eye, comprised quality and the volume of the speech during the conversations.[19,100,101]

Moreover, there are some environmental concerns associated with the use of single-use facemasks. Some of these facemasks are made from layers of plastics, which may not bio-degrade easily, thus creating a massive burden on the environment. A recent analysis has reported that if every person in the UK used one single-use facemask each day for a year, it would create 66,000 tonnes of contaminated plastic waste, roughly ten times higher than that of using reusable masks.

The new coronavirus is continuously evolving and spread all over the world. Researchers from all disciplines, especially the medical professionals and engineers, are continuously working on the facemasks design improvement for a better performance against the virus transmission. Zhou et. al.[102] presented an electrospun polyetherimide (PEI) electret nonwoven material based bi-functional smart facemask to remove the sub-micron particulate matter and generate electricity. The facemask could harvest sufficient energy from the airflow to supply power to the inbuilt LCD panel. The LCD screen was used to display the measured breathing rate. Hossain et al.[103] developed a rechargeable N95 facemask that composed of a charged polypropylene electret fiber made an intermediate layer for capturing the foreign particles. These particles are trapped through the electrostatic or electrophoretic effects of the polypropylene terephthalate (PET) layer. The mask has a provision for the in-situ recharging of the polypropylene electret for maintaining its filtration performance. Williams et al.[104] proposed a facemask used for the sample collection of respiratory SARS-CoV-2 virus. They have successfully presented a facemask prototype the detects exhaled Mycobacterium tuberculosis, a deadly lung infection, and now working for sampling for the SARS-CoV-2 virus. The facemask consisted of four 3D printed polyvinyl alcohol (PVA) sampling strips attached inside it. The sampling matrices trapped the particulates during exhalation and was further post-processed for the virus diagnosis. Face-mask sampling offered a highly efficient and non-invasive method for respiratory disease diagnosis. The presented approach showed great potential for diagnosis and screening, particularly in resource-limited settings.

Moreover, several innovative facemask prototypes with better filtration performance are available in the market. Recently, Korean electronics and appliance company LG® Ltd. has developed an air purifier Wearable mask (PuriCare™)[105] equipped with battery-operated miniature fans that draw in the fresh air and help reduce stuffiness. The Massachusetts Institute of Technology and Brigham and Women Hospital, Boston's researchers have developed the, a silicone-based transparent reusable facemask with a comparable performance level with N95 respirators.[106]

## VII. Summary

The facemasks have shown their potentials for preventing the spread of respiratory disease. A variety of facemasks ranging from a simple homemade cloth mask to the ventilated respirators, have played their role in the current COVID-19 pandemic. In general, the facemasks have been experimentally characterized by determining the filtration efficiency and total inward leakage ratio. Also, the fluid flow dynamics-based numerical methods have gained much attention to investigating the facemask performances. The present article has also highlighted the insufficiencies of assessing the breathing resistance of the wearers with the facemask by just examining the flow resistance of the facemask. In the longer term, there may be a need for a more elaborate system approach including the study and modeling of how the human lung would respond to the increase in breathing resistance due to the use of facemask, drawing the analogy of modeling the behavior of the heart for the blood circulation system. This article summarizes the perspective of the fluid dynamics of the facemask filtration performance, including droplet and aerosol transports, droplet evaporation, and facemask aerodynamics. Furthermore, recent investigations for the efficacy of the facemasks in the context of respiratory virus transmission have been discussed.

**Acknowledgement**
The first author would like to acknowledge the financial support from the Ministry of education RSB Research Fellowship Singapore.

**Data availability**
The data that support the findings of this study are available from the corresponding author upon reasonable request.